\documentclass[prl,superscriptaddress,showpacs,twocolumn]{revtex4}

\usepackage{graphicx}

\usepackage{graphicx,amsfonts}

\usepackage{epsfig,amsmath}

\usepackage{verbatim}

\begin{document}

\newcommand{\atanh}
{\operatorname{atanh}}

\newcommand{\ArcTan}
{\operatorname{ArcTan}}

\newcommand{\ArcCoth}
{\operatorname{ArcCoth}}

\newcommand{\Erf}
{\operatorname{Erf}}

\newcommand{\Erfi}
{\operatorname{Erfi}}

\newcommand{\Ei}
{\operatorname{Ei}}

\newcommand{\sgn}{{\mathrm{sgn}}}
\newcommand{\rme}{{\mathrm{e}}}
\newcommand{\rmd}{{\mathrm{d}}}
\def\be{\begin{equation}}
\def\ee{\end{equation}}

\def\bea{\begin{eqnarray}}
\def\eea{\end{eqnarray}}

\newcommand{\nn}{{\nonumber}}

\def\e{\epsilon}
\def\l{\lambda}
\def\d{\delta}
\def\o{\omega}
\def\cb{\bar{c}}
\def\Li{{\rm Li}}

\title{An exact solution for the KPZ equation with flat initial conditions}

\author{Pasquale Calabrese}
\affiliation{Dipartimento di Fisica dell'Universit\`a di Pisa and INFN, 56127 Pisa Italy}
\author{Pierre Le Doussal}
\affiliation{CNRS-Laboratoire de Physique
Th{\'e}orique de l'Ecole Normale Sup{\'e}rieure, 24 rue Lhomond, 75231
Paris Cedex, France}

\date{\today}

\begin{abstract}
We provide the first exact calculation of the height distribution at arbitrary time $t$ of the continuum KPZ growth equation
in one dimension with flat initial conditions. We use the mapping onto a directed polymer (DP) with one end fixed, one free,
and the Bethe Ansatz for the replicated attractive boson model. We obtain the generating function 
of the moments of the DP partition sum as a Fredholm Pfaffian. Our formula, valid for all times, exhibits convergence of the free energy (i.e. KPZ height) distribution to the GOE Tracy Widom distribution 
at large time. 
\end{abstract}

\maketitle

The continuum Kardar-Parisi-Zhang (KPZ) equation is the simplest equation describing the non-equilibrium
growth in time $t$ of an interface of height $h(x,t)$ in presence
of noise \cite{KPZ}. In one dimension, i.e. $x \in R$, it reads
\be \label{kpzeq}
\partial_t h = \nu \nabla^2 h + \frac{1}{2} \lambda_0 (\nabla h)^2 + \eta(x,t)\,,
\ee
where $\overline{\eta(x,t) \eta(x,t')}=D \delta(x-x') \delta(t-t')$ is a centered Gaussian white noise. 
Originally conceived to describe growth by random deposition and diffusion, it defines
a universality class believed to encompass an astounding variety of models and physical systems \cite{kpzreviews}.
The growing KPZ interface becomes, at large $t$, statistically self-affine with universal scaling exponents.
In $d=1$, its width is predicted \cite{exponent} to grow as $\delta h \sim t^{1/3}$, as observed in experiments \cite{exp1,exp2}. The KPZ problem also maps to forced Burgers turbulence \cite{Burgers}, 
and to the equilibrium statistical mechanics of a directed polymer (DP) in a random
potential, the simplest example of a glass \cite{directedpoly} with applications to 
vortex lines \cite{vortex}, domain walls \cite{lemerle}, and biophysics \cite{hwa}.
However, despite its importance and universality, the KPZ equation has
vigorously resisted analytical solutions.

Progress in analytical understanding of the KPZ class in $d=1$ came from
exact solutions of a lattice DP model at zero temperature \cite{Johansson2000},
discrete growth models such as the PNG model \cite{spohn2000,png}, 
asymmetric exclusion models \cite{spohnTASEP} and vicious 
walkers \cite{greg}. An analogous to the
height field $h(x,t)$ was identified and, in the large size limit, its one-point (scaled) probability
distribution was shown to equal the (scaled) distribution of the smallest eigenvalue of a 
random matrix drawn from the famous Gaussian ensembles, the so-called
Tracy Widom (TW) distribution \cite{TW1994}, which appears in many
other contexts \cite{othersTW}.
It was found \cite{spohn2000,ferrari1} that 
one gets either the TW distribution $F_2(s)$ of the Gaussian unitary ensemble (GUE)  or $F_1(s)$
of the Gaussian orthogonal  ensemble (GOE) for droplet and flat {\it initial conditions} respectively. 
The corresponding many point distributions were identified as 
determinantal space-time processes, the Airy process, $Ai_2$ for droplet, and $Ai_1$ for flat
naturally expressed with the use of Fredholm determinants \cite{spohnreview}.

These advances gave valuable, but only indirect information on the {\it continuum}
KPZ equation, i.e. a conjecture for its infinite $t$ limit (termed the KPZ renormalization fixed
point \cite{corwinRG}). Only recently we \cite{we}, and other workers \cite{dotsenko,spohnKPZEdge,corwinDP,airynew}, were able
to directly solve the continuum problem and, until now, only within the droplet initial condition.
In addition to the convergence to $F_2(s)$ at large $t$,  the unveiled
remarkable feature  is that a proper generating function 
$g(s)$ (recalled below) remains a Fredholm determinant {\it for all times $t$}, hence
leading to an exact solution for the universal crossover in time in the continuum KPZ equation.
This universal distribution (depending on a single parameter $t$) describes in the DP framework the high temperature regime \cite{we}
that has remarkable universal features \cite{highT}. In the growth problem,
this corresponds to a universal large diffusivity-weak noise limit, at fixed 
correlation length of the noise.

In this Letter we obtain the corresponding exact result for the continuum KPZ equation 
for the case of {\it flat} initial conditions, most often encountered in experiments
\cite{exp2}. We obtain here the generating function of the integer moments of the DP partition sum 
$Z\equiv e^{\frac{\lambda_0}{2 \nu} h}$ (see below)
\be \label{gdef1}
g_\lambda(s) = \sum_{n=0}^\infty \frac{(- e^{- \lambda s})^n}{n!} \overline{Z^n}  \quad , \quad 
\lambda = \frac{1}{2} (\bar c^2 t/T^5)^{1/3}\,,
\ee
with $\bar c=D \lambda_0^2$, $T=2 \nu$, as a Fredholm Pfaffian for any time $t$. The DP free energy, and
the height field at a given point, take the form
\bea
\frac{\lambda_0}{2 \nu} h \equiv \ln Z = v_0 t + \lambda \xi_t \,,
\eea 
where $g_\lambda(s) = \overline{ \exp( - e^{\lambda (\xi_t-s)} ) }$. In the large time
limit $g_\infty(s)\equiv \lim_{\lambda \to \infty} g_\lambda(s) = {\rm Prob}(\xi_t < s)$ and we find
that the distribution variable $\xi_t$ converge to  the GOE 
Tracy Widom distribution ${\rm Prob}(\xi_t < s) = F_1(s)$. As in \cite{we} we use the Bethe Ansatz (BA)
for the replicated boson model with $\delta$ attraction, and sum over all excited states,
treating now the case of a DP with one end fixed and the other free. The calculation is far
more complicated than \cite{we} as we need the spatial integrals of the Bethe wavefunctions.
Technically, this is surmounted by solving a half-space model in the proper limit. 

The solution of (\ref{kpzeq}) for a given initial condition can be written, using the Cole-Hopf 
transformation $Z\equiv e^{\frac{\lambda_0}{2 \nu} h}$
\be \label{CO}
e^{\frac{\lambda_0}{2 \nu} h(x,t)} = \int dy Z(x,t|y,0) e^{\frac{\lambda_0}{2 \nu} h(y,t=0)}\,,
\ee
in terms of the partition function of a DP, at temperature $T=2 \nu$ and in 
the random potential $V(x,t)=\lambda_0 \eta(x,t)$, i.e the sum over paths $x(\tau) \in R$ starting at $x(0)=y$ and ending at $x(t)=x$
\be \label{zdef} 
Z(x,t|y,0) = \int_{x(0)=y}^{x(t)=x}  Dx(\tau) e^{- \frac{1}{T} \int_0^t d\tau [ \frac{1}{2}  (\frac{d x}{d\tau})^2  + V(x(\tau),\tau) ]} \,,
\ee
with initial condition $Z(x,t=0|y,0)=\delta(x-y)$. In this continuum model the disorder
correlation length is zero, i.e. 
$\overline{V(x,t) V(x',t)} = \bar c \delta(t-t') \delta(x-x')$ with $\bar c = D \lambda_{0}^2$,
hence one can perform a rescaling $x \to T^3 x/\bar c$ and $t \to 2 T^5 t/\bar c^2$, and
work in units such that $T=1$ and $\bar c=1$, as we do below. 
The continuum model (\ref{zdef}) describes the high $T$ limit of the DP on a lattice
as discussed in \cite{highT,we}. For the KPZ interface
this continuum model describes the universal limit where 
the characteristic time $t^* = 2 (2 \nu)^5/D^2 \lambda_0^4$ and space $x^*=\sqrt{\nu t^*}$ scales
are much larger than the correlation lengths of the noise.

In \cite{we} we obtained the distribution of $\ln Z$ for a DP with the two ends fixed $Z=Z(0,t|0,0)$.
From (\ref{CO}) this corresponds to a wedge initial condition for the KPZ interface, $\frac{\lambda_0}{2 \nu} h_{\rm wedge}(x,t=0)= - w |x|$, in the limit
of a narrow wedge $w \to \infty$. Here we solve the {\it flat interface} initial conditions for KPZ, i.e. $h(x,t=0)=0$,
hence the opposite limit $w \to 0^+$ of the wedge. It corresponds to the DP with one end fixed, one end free, i.e. 
$Z_{\rm flat}(x,t) = \int dy Z(x,t|y,0)$. To achieve that we found it easier to study the (left) half-space
problem
\begin{eqnarray}
&& Z_w(x,t) = \int_{-\infty}^{0} dy e^{w y} Z(x,t|y,0)\,,
\eea
with $Z(x,t=0)=\theta(-x) e^{w x}$. It is easy to show \cite{uslong}
that at $w=0^+$ the half space model interpolates between (i) the 
narrow wedge initial condition for $x \to +\infty$, since the polymer is stretched
and (ii) the flat (full-space) initial condition for $x \to -\infty$, hence we
study below $\lim_{x \to - \infty} \lim_{w \to 0} Z_w(x,t)
=Z_{\rm flat}(x,t)$.

As well known \cite{kardareplica} the calculation of the $n$-th integer moment of 
a DP partition sum can be expressed as a quantum mechanical problem
for $n$ particles described by the (attractive) Lieb-Liniger Hamiltonian \cite{ll}
\be
H_n = -\sum_{j=1}^n \frac{\partial^2}{\partial {x_j^2}} 
- 2 \bar c \sum_{1 \leq  i<j \leq n} \delta(x_i - x_j).
\label{LL}
\ee
Generalizing \cite{we}, the quantum mechanical expectation for $ \overline{Z_w(x,t)^n}$ is written as a sum over the 
un-normalized eigenfunctions $\Psi_\mu$ (of norm denoted $||\mu ||$) of $H_{n}$ with energies $E_\mu$
\footnote{
For convenience we exchanged the DP endpoints, using that $Z_w$ is real and $H_n$ hermitian}:
\bea
&& \overline{Z_w(x,t)^n} =   \int_{y_i<0} e^{w \sum_{i=1}^n y_i } \langle y_1 \dots y_n |e^{- t H_n} |x \dots x \rangle \nn
\\
&& = \sum_\mu \Psi_\mu^*(x,..x)  \int^w \Psi_\mu
\frac{1}{||\mu ||^2} e^{-t E_\mu} \label{sum}\,, \\
&& \int^w \Psi_\mu := \int_{y_i<0} e^{w \sum_{i=1}^n y_i}  \Psi_\mu(y_1,..y_n)\,,  \label{int}
\end{eqnarray} 
where we used the fact that only symmetric (i.e. bosonic) eigenstates contribute. 
The Bethe states  $\Psi_\mu$ are superpositions of plane waves \cite{ll}  over all permutations $P$ of the rapidities 
$\l_j$ ($j=1,..n$) and we use the convention
\bea \label{bethe}
\Psi_\mu(x_1,..x_n) = \sum_P A_P \prod_{j=1}^n e^{i \l_{P_\ell} x_\ell} \,,
\eea
where the coefficients $A_P=\prod_{n \geq \ell > k \geq 1} (1+ \frac{i \bar c ~\text{sgn}(x_\ell - x_k))}{\lambda_{P_\ell} - \lambda_{P_k}})$.
The general eigenstates are built by partitioning the $n$ particles into a set of $n_s$ 
bound-states formed by $m_j \geq 1$ particles with $n=\sum_{j=1}^{n_s} m_j$.
Because we work with $w=0^+$, we can take directly the system size $L=\infty$
and in that limit \cite{m-65} each bound state is a {\it perfect string}, i.e. a set of
rapidities $\l^{j, a}=k_j +\frac{i\cb}2(m_j+1-2a)$, where $a = 1,...,m_j$ 
labels the rapidities within the string. Such eigenstates have momentum 
$K_\mu=\sum_{j=1}^{n_s} m_j k_j$ 
and energy $E_\mu=\sum_{j=1}^{n_s} (m_j k_j^2-\frac{\cb^2}{12} m_j(m_j^2-1))$. 
The ground-state corresponds to a single $n$-string with $k_1=0$.

In (\ref{sum}) one already knows $\Psi_\mu^*(x,..x)= n! e^{- i x \sum_\alpha \lambda_\alpha}$
and the norms $||\mu||$ of the string states  \cite{cc-07}
\begin{eqnarray}
&& ||\mu||^{-2} = \frac{(\bar{c})^{n}}{n!  (L \bar{c})^{n_s} }
\prod_{j=1}^{n_s} m_j^{-2} \prod_{\substack{1\leq i<j\leq n_s}}  \Phi_{k_i,m_i,k_j,m_j}\,, \nn \\
&& \Phi_{k_i,m_i,k_j,m_j}  :=
\frac{(k_i-k_j)^2 +(m_i-m_j)^2 \bar c^2/4}{(k_i-k_j)^2 +(m_i+m_j)^2 \bar c^2/4} \,.
\end{eqnarray}
The new difficulty, i.e. computing the spatial integral of the Bethe states,
simplifies dramatically for the half-space model. Using the symmetry
of Eq. (\ref{bethe}), we have
\begin{eqnarray}
&& \int^w  \Psi_\mu  = n! \sum_P G^w_{\lambda_{P_1},..\lambda_{P_n}} 
\prod_{n \geq \ell > k \geq 1} (1+ \frac{i \bar c}{\lambda_{P_\ell} - \lambda_{P_k}})\,, \nn \\
&& G^w_{\lambda_1,..\lambda_n} = \prod_{j=1}^{n} \frac{1}{ j w + i \lambda_1 + .. + i \lambda_j }\,.
 \nonumber 
\end{eqnarray}
From the remarkable properties of the BA, it can be reexpressed, for any $n$ and set of rapidities (with $\bar c=1$):
\be
\int^w \Psi_\mu   =  \frac{n!}{i^n \prod_{\alpha=1}^n (\lambda_\alpha - i w)}  \prod_{1 \leq \alpha < \beta \leq n} \frac{i + \lambda_\alpha + \lambda_\beta - 2 i w}{ \lambda_\alpha + \lambda_\beta - 2 i w} .
\ee
If we now inject the string solution $\lambda_{j,a}=\frac{i}{2} (m_j+1 - 2 a) + k_j$, we find after some
elementary manipulations
\bea
&& \int^w \Psi_\mu  = n! (-2)^{n} \prod_{i=1}^{n_s} S^w_{m_i,k_i}
 \prod_{1 \leq i < j \leq n_s} D^w_{m_i,k_i,m_j,k_j}\,, \nn \\
&& S^w_{m_i,k_i} = \frac{\Gamma (\kappa_{ii}-m_i)}{\Gamma (\kappa_{ii})} \,,   \\
&& D^w_{m_i,k_i,m_j,k_j} =
\frac{\Gamma \left(\kappa_{ij} -\frac{m_i+m_j}{2}\right) \Gamma \left(\kappa_{ij} +\frac{m_i+m_j}{2}\right) }{
\Gamma \left(\kappa_{ij} +\frac{m_i-m_j}{2}\right) \Gamma \left(\kappa_{ij} -\frac{m_i-m_j}{2}\right)}\,, \nn
 \eea
with $\kappa_{ij}=-i k_i - i k_j - 2 w + 1$. 

We have now all ingredients to compute the generating function (\ref{gdef1})
with $Z \equiv Z_w(x,t)$. Writing the sum over states in (\ref{sum}) as all partitioning of 
$n$ particles into $n_s$ strings and using that for $L\to\infty$ the string
momenta $m_j k_j$ correspond to free particles \cite{cc-07} (i.e. 
$\sum_{k_j} \to m_j L \int \frac{dk_j}{2 \pi}\equiv m_j L \int _{k_j}$), Eq. (\ref{gdef1}) becomes a sum over string 
configurations $g_\lambda(s) = 1 +  \sum_{n_s=1}^\infty \frac1{n_s!}  Z(n_s,s)$ with
\begin{widetext}
\be
 Z(n_s,x) =  \sum_{m_1,\dots m_{n_s}=1}^\infty 
 \prod_{j=1}^{n_s} [ \frac{2^{m_j}}{m_j} \int_{k_j} 
S^w_{m_j,k_j}  e^{(m_j^3 - m_j) \frac{t}{12}- m_j k_j^2 t - \lambda m_j s - i x m_j k_j} ]  \prod_{1 \leq i < j \leq n_s} D^w_{m_i,k_i,m_j,k_j}  \Phi_{k_i,m_i,k_j,m_j} \,.
\ee
\end{widetext}

Upon inspection we find that the limit of interest, $w=0^+$, is dominated by poles in the $k_i$ integrations, with $S^w_{m_i,k_i} \sim \frac{(-1)^{m_i}}{\Gamma(m_i) (2 i k_i + 2 w)}$ and
$D^w_{m_i,k_i,m_j,k_j} \sim \frac{(-1)^{m_i} m_i}{ i (k_i +k_j)+ 2 w} \delta_{m_i,m_j}$, and that the regular parts do not
contribute. Replacing $1/(i k + 0^+) \to \pi \delta(k)$ yields an $x$-independent result, which can be
shown to equal the limit $x \to -\infty$, i.e. the flat initial condition for KPZ, on which we now focus.
The result is the sum of the residues associated to configurations where
the $n_s=2 N+M$ strings split into $N$ pairs of strings of opposite momenta with same particle
number $m$ and $M$ single strings of zero momentum with all distinct number of particles. 
After some non-trivial manipulations, detailed in \cite{uslong}, we bring the result in the form
of a Pfaffian:
\begin{widetext}
\bea &&
Z(n_s) = \sum_{m_i \geq 1}  
   \prod_{j=1}^{n_s} \int_{k_j} \prod_{q=1}^{m_j} \frac{-2}{2 i k_j + q} 
   e^{\frac{\lambda^3}{3} m_j^3 - 4 m_j k_j^2 \lambda^3 - \lambda m_j s} 
\\ &&
\times 
{\rm Pf}\bigg[
\left(\begin{array}{cc}
 \frac{2 \pi}{2 i k_i} \delta(k_i+k_j) (-1)^{m_i} \delta_{m_i,m_j} + \frac{1}{4} (2 \pi)^2 \delta(k_i) \delta(k_j) (-1)^{\min(m_i,m_j)} {\rm sgn}(m_i-m_j) & \frac{1}{2}  (2 \pi) \delta(k_i)   \\
- \frac{1}{2} (2 \pi)  \delta(k_j)  &  \frac{2 i k_i + m_i - 2 i k_j - m_j}{2 i k_i + m_i + 2 i k_j + m_j} 
\end{array} \right)\bigg]_{2 n_s  \times 2 n_s} . \nonumber 
\eea
\end{widetext}
We recall that for an antisymmetric matrix $A$ of size $2 n_s$
\be
{\rm Pf} A = \sum_{\sigma \in S_{2n_s}, \sigma(2j-1)<\sigma(2 j)} \hspace{-2mm} (-1)^\sigma \prod_{i=1}^{n_s} 
A_{\sigma(2 i-1),\sigma(2 i)}\,,
\ee
with $({\rm Pf} A)^2 = {\rm det} A$. We can now use the Airy trick \cite{dotsenko,we}
$\prod_j e^{\frac{1}{3} \lambda^3 m_j^3}=\prod_j \int_{y_j} Ai(y_j) e^{\lambda y_j}$ and
decouple the denominators in the lower right corner using
auxiliary integrals $\prod_j \int_{v_j>0} e^{-v_j A_j} = \prod_j \frac{1}{A_j}$ 
and the numerators using derivatives. We perform rescaling $k_j \to k_j/\lambda$, and
shifts $y_j \to y_j + v_j - 4 k_j^2 + s$. The summations over the $m_i$ can
be performed exactly inside the Pfaffian and we arrive at our main result for
$g_\lambda(s)$ as a {\it Fredholm Pfaffian}
\bea
&& g_\lambda(s) = {\rm Pf}[ {\bf J} + {\bf K} ] = \sum_{n_s=0}^\infty \frac{1}{n_s!} Z(n_s)\,, \\
&& Z(n_s) = \prod_{j=1}^{n_s} \int_{v_j>0} {\rm Pf}[{\bf K} (v_i,v_j)]_{2n_s,2n_s}\,, \nn
\eea
where ${\bf J}=
\left(\begin{array}{cc}
0 & I  \\
- I & 0
\end{array} \right)$ and
${\bf K}$ is an antisymmetric $2$ by $2$ matrix kernel of components $K_{ab} \equiv K_{ab}(v_i,v_j)$ with
\begin{widetext}
\bea
&&  K_{11} = \int_{y_1,y_2,k} Ai(y_1+v_i+s+4k^2)  Ai(y_2+v_j+s+4k^2) [
  \frac{e^{- 2 i (v_i -v_j) k}}{2 i k}  f_{k/\lambda}(e^{ \lambda (y_1+y_2)}) + \frac{\pi \delta(k) }{2} 
  F(2 e^{\lambda y_1} , 2 e^{\lambda y_2}) ]\,, \nn 
  \eea
  \end{widetext}
\bea 
  &&    K_{12}= \frac{1}{2} \int_{y} Ai(y+s+v_i) (e^{-2 e^{\lambda y}}-1)  ~ \delta(v_j)\,,  \nn \\
  &&  K_{22} = 2\delta'(v_i-v_j) \,,
  \eea
and the functions 
  \bea
  && f_k(z)  = \frac{- 2 \pi  k  z \, _1F_2\left(1;2-2 i k,2+2 i
   k;- z\right)}{\sinh \left(2 \pi  k \right) \Gamma \left(2- 2 i k \right)
   \Gamma \left(2+2 i k\right)}\,, \\
   && F(z_i,z_j)=  \sinh(z_2-z_1) +e^{-z_2} -e^{-z_1}  + 
\int_0^{1} du  \nn \\
&& \times J_0(2 \sqrt{z_1 z_2 (1-u)}) [ z_1 \sinh(z_1u) - z_2 \sinh(z_2 u)]\,. \nn
\eea
The full analysis of this result is performed in \cite{uslong}. Here, we
first point out the simple one-string contribution ($n_s=1$), 
$Z(1)  = \int_{v>0}  K_{12}(v,v)$ leading to
\be \label{1}
Z(1) = \frac{1}{2} \int dy  (e^{-2 e^{\lambda y}}-1) Ai(y+s) \,,
\ee
also obtained \cite{uslong} from the ground state for each $n$, 
which gives the leading asymptotics of $g_\lambda(s)$ for large $s>0$. 
Using $\det \left[\begin{array}{cc} A& B\\ C &D\end{array}\right]=\det D \det [A -BD^{-1} C]$, $g_\lambda^2(s)= 
{\rm Det}[{\bf I} - {\bf J} {\bf K}]$ can be written 
(for any time) in a form suitable for numerical evaluation \cite{uslong}.
The Pfaffian reported above allows simpler analytic manipulations. 

In the large time (large $\lambda$) limit, one already sees from (\ref{1}) that 
$Z(1) \to - \int_{y>0} Ai(2 y + s) = - Tr {\cal B}_s$
where ${\cal B}_s = \theta(x) Ai(x+y+s) \theta(y)$ is the GOE kernel, as
shown by Ferrari and Spohn \cite{ferrari2}. This extends to all $n_s$ i.e. we find that
\be \label{lim}
\lim_{\lambda \to + \infty} Z(n_s)=(-1)^{n_s} \int_{x_1,..x_{n_s}} \hspace{-3mm} \det[{\cal B}_s(x_i,x_j)]_{n_s \times n_s}. 
\ee 
Hence $g_\infty(s) = F_1(s) = \det[I - {\cal B}_s]$ 
the Fredholm determinant expression for the GOE Tracy Widom distribution. 
This is obtained using $\lim_{\lambda \to + \infty} f_{k/\lambda}(e^{ \lambda y})
= - \theta(y)$ and $\lim_{\lambda \to + \infty} F(2 e^{\lambda y_1} , 2 e^{\lambda y_2}) 
= \theta(y_1+y_2) (\theta(y_1)\theta(-y_2) - \theta(y_2)\theta(-y_1))$. 
We checked (\ref{lim}) explicitly up to $n_s=4$, but we report the proof in
\cite{uslong}. For $n_s$ even it follows from a slight generalization of \cite{ferrari2}, namely
${\rm det}(I \mp {\cal B}_s)/{\rm det}(I \pm  {\cal B}_s)=\int_{x>0} (I \pm {\cal B}_s)^{-1}(x,0)$. 


To summarize we have obtained the generating function for the distribution of the free energy 
of the DP with one free end, i.e. of the height of the continuum KPZ interface, for
arbitrary time. At large time the distribution crosses over to the GOE Tracy Widom distribution
$F_1(s)$. Further properties of the finite time, including extracting $P(f)$ and numerics
are studied in \cite{uslong}. 

We thank A. Rosso for discussions and for helpful numerical checks \cite{uslong} of (i) low integer moments of $Z$ 
at small $t$ (ii) the variance of $\ln Z$ at large $t$.
PC thanks LPTENS, and PLD thanks KITP for
hospitality. This work was supported by ANR grant 09-BLAN-0097-01/2.

\end{document}